# Dynamic Equivalence between Soft Star Polymers and Hard Spheres

Zhe Wang[†,‡,*], Antonio Faraone[§], Panchao Yin[‖], Lionel Porcar[⊥], Yun Liu[§], Changwoo Do[‡], Kunlun Hong[¶], Wei-Ren Chen[‡,*]

[†]*Department of Engineering Physics & Key Laboratory of Particle & Radiation Imaging (Tsinghua University) of Ministry of Education, Tsinghua University, Beijing 100084, China*

[‡]*Neutron Scattering Division, Oak Ridge National Laboratory, Oak Ridge, TN 37831, USA*

[§]*Center for Neutron Research, National Institute of Standards and Technology, Gaithersburg, MD 20899-6100, USA*

[‖]*South China Advanced Institute for Soft Matter Science and Technology, South China University of Technology, Guangzhou 510640, China*

[⊥]*Institut Laue-Langevin, B.P. 156, F-38042 Grenoble CEDEX 9, France*

[¶]*Center for Nanophase Materials Sciences, Oak Ridge National Laboratory, Oak Ridge, TN 37831, USA*

[*]To whom correspondence should be addressed. E-mail: zwang2017@mail.tsinghua.edu.cn (Z.W.); chenw@ornl.gov (W.-R. C.)

**TOC Graphic**

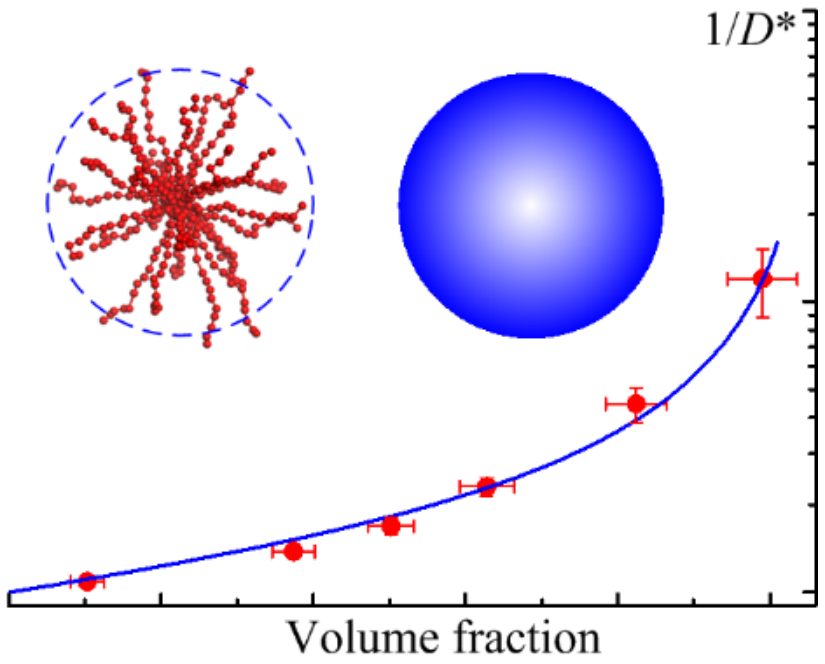

**Abstract**

Understanding the dynamics of soft colloids, such as star polymers, dendrimers, and microgels, is of scientific and practical importance. It is known that the excluded volume effect plays a key role in colloidal dynamics. Here, we propose a condition of *compressibility equivalence* that provides a simple method to experimentally evaluate the excluded volume of soft colloids from a thermodynamic view. We apply this condition to survey the dynamics of a series of star polymer dispersions. It is found that as the concentration increases, the slowing of the long-time self-diffusivity of the star polymer, normalized by the short-time self-diffusivity, can be mapped onto the hard-sphere behavior. This phenomenon reveals the dynamic equivalence between soft colloids and hard spheres, despite the apparent complexity of the interparticle interaction of the soft colloids. The methods for measuring the osmotic compressibility and the self-diffusivities of soft colloidal dispersions are also presented.

Dispersions of soft deformable colloids, such as star polymers, dendrimers, microgels, and polymer-grafted nanoparticles, constitute a great portion of soft matters.[1-4] These materials exist extensively in the fields of food, pharmaceutical, and chemical engineering. Therefore, measuring and modeling their physical properties are of practical importance. A fundamental challenge in the physics of soft colloids is to quantitatively predict the slowing of their long-time dynamics as the concentration increases, since this behavior is closely related to many essential phenomena, for example, the phase behavior, glass formation, and rheology.[3-8] Over past decades, several theoretical and computational efforts have been made on this topic.[9-14] In these studies, researchers introduced the effective volume fraction $\phi_{\mathrm{eff}}$ (or effective diameter $d_{\mathrm{eff}} = (6\phi_{\mathrm{eff}}/n\pi)^{1/3}$ where $n$ is the number density of particles) for soft particles due to their excluded volume effect at high concentrations. With $\phi_{\mathrm{eff}}$, the long-time dynamics of soft particles can be scaled to the hard-sphere behavior,[10-13] which is well investigated both experimentally and theoretically.[15] The key prerequisite for these scaling operations is the pair potential $v(r)$[10] ($r$ is the distance between two colloids) or the interparticle structure factor $S(Q)$[11-13] ($Q$ is the magnitude of the scattering vector in scattering experiments) of the soft colloids. However, without knowing the exact way how the particle deforms as concentration increases, and an accurate functional form of the pair potential at high concentrations, it is very difficult to measure $v(r)$ and $S(Q)$ precisely from scattering experiments, especially at high concentrations. To address this problem from an experimental perspective, we propose a condition of *compressibility equivalence* that provides a remarkably simple way to determine the effective volume fraction of soft colloids from the small-angle neutron scattering (SANS) experiment. Employing this condition, we find that the slowing of the long-time self-diffusion of soft colloids can be nicely mapped onto the property of hard spheres.

A series of star polymer dispersions are measured to study the dynamics of colloidal liquids with soft repulsive interactions. Star polymers are synthetic macromolecules consisting of polymeric branches emanating from the molecular center. Experimental and theoretical studies show that star polymers exhibit colloidal nature in dispersions.[2,16-20] Due to their flexible molecular architecture, the effective potential between two stars has been modeled as an *ultrasoft* repulsion.[21,22] In this work, 15-arm polystyrene stars, with on average 12.5 monomers contained in each arm, are used as the model soft deformable colloids. The star polymer samples were purchased from Polymer Source, Inc.. The molecular weight of one star, determined from the light scattering, is $1.9 \times 10^4$ g/mol. Cyclohexanone is adopted as the good solvent. 6 concentrations of stars, 3, 10, 15, 20, 30, and 40 wt.% (wt.% indicates the weight fraction in percent), are investigated. Figure 1 displays the SANS spectra of the protonated stars immersed in fully deuterated solvents at the above 6 concentrations. Upon increasing the concentration, the interaction peak grows gradually up to about 15 wt.%, and then decays with further concentration increment. This non-monotonic evolution of the main peak has been identified as the generic feature of soft colloidal systems[2,23] and its origin has been attributed to the diminishing density fluctuation at molecular level due to the increasing mutual interaction.[21,24] Notice that, this behavior is significantly different from the behaviors of hard spheres, where the interaction peak progressively grows as concentration increases.[15]

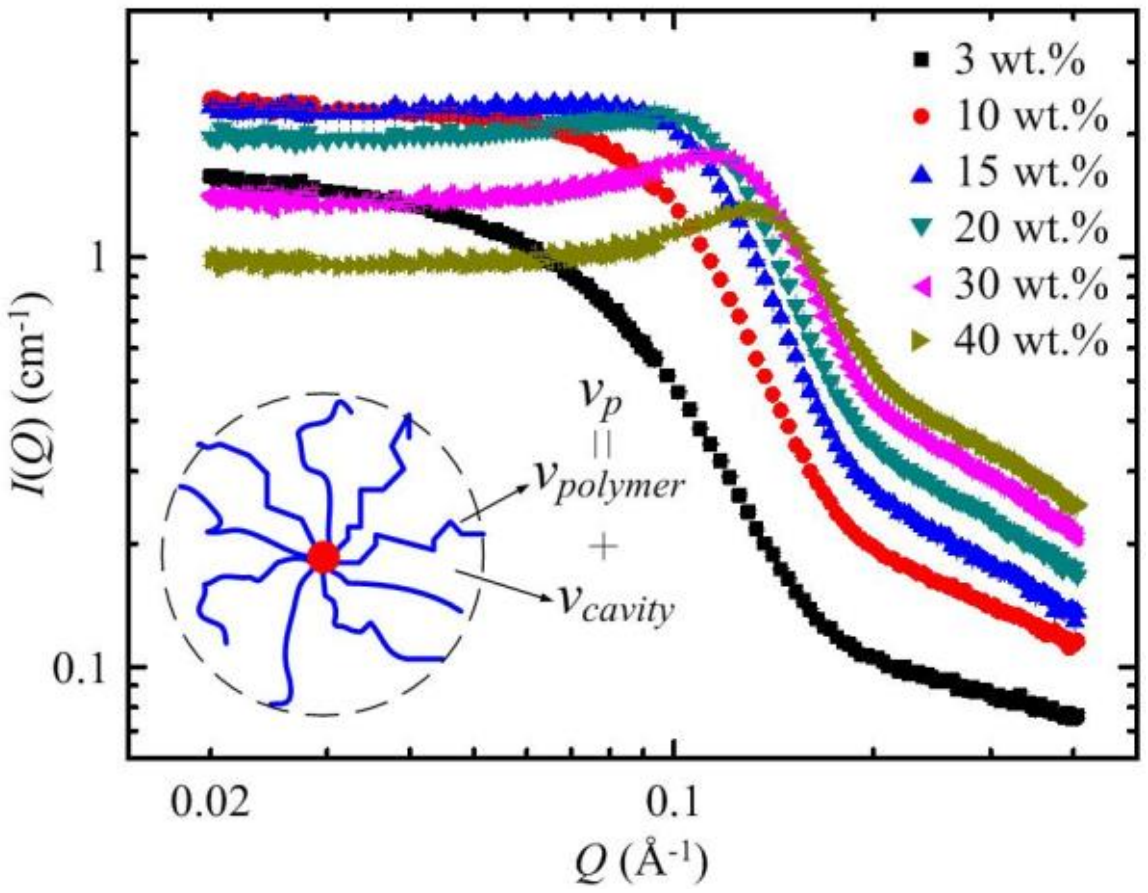

**Figure 1.** The SANS spectra of the star polymer dispersions at 6 measured concentrations. The inset illustrates the meaning of $v_{\mathrm{p}}$, $v_{\mathrm{polymer}}$ and $v_{\mathrm{cavity}}$ in a star polymer dispersion.

It is instructive to briefly review a few theoretical results before further analysis. There are two well-defined self-diffusive processes in colloidal suspensions: the short-time self-diffusivity $D_{\mathrm{SS}}$ and the long-time self-diffusivity $D_{\mathrm{LS}}$.[5,8,25-28] $D_{\mathrm{SS}}$ describes the diffusion of the particle within its neighbor cage, while $D_{\mathrm{LS}}$ describes the self-diffusion on a time scale longer than that of the local structural rearrangement.[25,26] At the dilute limit, both of them equal to the Stokes-Einstein diffusivity of an isolated particle: $D_0 = kT/3\pi\eta_s a$, where $k$ is the Boltzmann constant, $T$ the temperature, $\eta_{\mathrm{s}}$ the solvent viscosity, and $a$ the hydrodynamic diameter of the particle. As concentration increases, $D_{\mathrm{SS}}$ is retarded due to the emergence of the hydrodynamic interaction.[5,27] In the long-time self-diffusion, the tagged particle breaks the neighbor cage and moves to relatively distant places. This process requires the distortion and relaxation of the local configuration.[25,26] As concentration increases, it slows down more significantly than the short-time self-diffusion due to the excluded volume effect of particles. Considering that the mechanisms of the slowing of $D_{\mathrm{SS}}$ and $D_{\mathrm{LS}}$ are distinct, and the time scales of $D_{\mathrm{SS}}$ and $D_{\mathrm{LS}}$ are well separated at high concentrations, theorists point out that $D_{\mathrm{LS}}$ can be approximately factorized into a hydrodynamic part and a thermodynamic or structural part, namely,[25,26]

$$D_{\mathrm{LS}} \approx D_{\mathrm{SS}} f\big(S(Q)\big), \quad (1)$$

where $f\big(S(Q)\big)$ is a factor depending on the properties of $S(Q)$ (or the pair distribution function $g(r)$). Therefore, it is natural to introduce a normalized long-time self-diffusivity $D^* = D_{\mathrm{LS}}/D_{\mathrm{SS}}$ to characterize the long-time dynamics of the star polymers. $D_{\mathrm{SS}}$, on the other hand, plays a role as a coarse-grained time resolution. We would like to point out that the influence of hydrodynamic interaction on $D_{\mathrm{LS}}$ is more complicated than the decoupling approximation given by eq 1.[28-31] A recent study shows that in charged colloidal suspensions, the hydrodynamic interaction enhances (rather than slowing down) the long-time self-diffusion by about 15% at certain concentrations.[31] Nevertheless, we argue that $D^*$ is a proper quantity that reflects the thermodynamic or structural basis of the long-time self-diffusion of colloids, by considering that (i) the drastic slowing of $D_{\mathrm{LS}}$ as concentration increases is mainly due to the structure rather than the hydrodynamic interaction, and (ii) the approximation given by eq 1 explains experimental data well, as suggested in References 25 and 26. In following analyses, we will use $D^*$ as a measure of the long-time dynamics of the star polymers.

To describe the dynamics of soft particles with hard-sphere properties, it is crucial to find the effective volume fraction $\phi_{\mathrm{eff}}$ for soft particles based on certain structural or thermodynamic similarities between these two systems. Inspired by previous theoretical results,[9-13] we propose that, for a soft repulsive colloidal dispersion with the number density of colloids $n$, $\phi_{\mathrm{eff}}$ is determined by the following condition:

$$\lim_{Q\to 0} S_{\mathrm{sc}}(Q;n) = \lim_{Q\to 0} S_{\mathrm{hs}}(Q;n,\phi_{\mathrm{eff}}), \quad (2)$$

where $S_{\mathrm{sc}}(Q;n)$ is the interparticle structure factor of the soft colloidal dispersion with number density of $n$, $S_{\mathrm{hs}}(Q;n,\phi_{\mathrm{eff}})$ is the structure factor of the monodisperse hard-sphere system with number density of $n$ and volume fraction of $\phi_{\mathrm{eff}}$. Since $S(Q\to 0)$ is related to the osmotic compressibility $\chi$ of colloidal dispersions by $S(Q\to 0) = nkT\chi$,[15,32] we call eq 2 as the condition of compressibility equivalence. This equivalence suggests that the two systems bear a resemblance to each other regarding the effect of excluded volume of all particles. In fact, for liquids with purely repulsive potential, $\chi$ is largely determined by the second virial coefficient that gives the excluded volume of a particle in the van der Waals picture. In eq 2, $S_{\mathrm{hs}}(Q;n,\phi_{\mathrm{eff}})$ can be calculated with Percus-Yevick approximation[33] and Verlet-Weis correction.[34] Thus, $\phi_{\mathrm{eff}}$ can be obtained when $S_{\mathrm{sc}}(Q\to 0;n)$ is known. In the following part, we will show that with $\phi_{\mathrm{eff}}$ defined by eq 2, the $D^*$ of star polymers is found to be equivalent to that of the hard spheres.

The theoretical foundation of eq 2 is close to the scheme proposed by Xu *et al.*.[9] With computer simulation, these authors investigated many soft-core liquids. Some of them are with very soft interactions, such as harmonic repulsion. They found that at low pressures, the dynamics of soft-core liquids and hard-sphere liquids can be collapsed to a single curve as a function of $T/Pd^3$, where $P$ is pressure and $d$ is the diameter of the soft repulsion. By employing eq 2 at all concentrations, we ensure that the soft particles and hard spheres have a common $\mathrm{d}P/\mathrm{d}\phi_{\mathrm{eff}}$ in the case that $d_{\mathrm{eff}}$ does not change too much as concentration increases. Consequently, the soft particles and hard spheres are with similar equations of state, *i.e.*, the pressure as a function of volume fraction. Therefore, the condition of compressibility equivalence leads to similar dynamical properties of soft particles and hard spheres according to the scheme given by Xu *et al*.

Nevertheless, this dynamic scaling may fail at very high concentrations. For typical colloidal systems, a volume fraction higher than about 50% may violate the low-pressure requirement of this scaling law.

Another scheme, proposed by Medina-Noyola *et al.*, states that the $\phi_{\mathrm{eff}}$ determined by the following condition maps the long-time self-diffusion of soft colloids onto that of hard spheres:[11-13]

$$S_{\mathrm{sc}}(Q;n) \approx S_{\mathrm{hs}}(Q;n,\phi_{\mathrm{eff}}). \quad (3)$$

The symbol "≈" in preceding equation means that the two structure factors match at low and intermediate $Q$, corresponding to the main peak of $S(Q)$.[11-13,35,36] In real space, eq 3 indicates that the pair distribution functions $g(r)$ of the two systems are approximately equal at $r > \sim 1.3 d_{\mathrm{eff}}$.[15,36] It emphasizes the similarity of the overall nature of packing, rather than the details at the contact, of the two systems. Therefore, we arguably state that the physical backgrounds of equations 2 and 3 are consistent. In fact, the Brownian dynamics data given in Reference 13 suggest that both equations 2 and 3 can scale the data of $D^*$ of colloids with investigated soft repulsions to the hard-sphere behavior.

To experimentally measure $D^*$ of soft colloids and verify its possible equivalence to the hard-sphere behavior, one needs to probe $S_{\mathrm{sc}}(Q \to 0)$, $D_{\mathrm{SS}}$ and $D_{\mathrm{LS}}$. We devise a model-independent method using the contrast-variation SANS to find $S_{\mathrm{sc}}(Q \to 0)$. For a monodisperse colloidal dispersion,[37] the SANS spectrum is expressed as:[38]

$$I(Q) = n v_{\mathrm{p}}^2 (\Delta\rho^{\mathrm{sld}})^2 P(Q) S(Q), \quad (4)$$

where $n$ is the number density of colloid particles, $v_{\mathrm{p}}$ is the volume occupied by one colloidal particle, $\Delta\rho^{\mathrm{sld}}$ is the contrast of the scattering length density (sld) between the colloidal particle and the solvent, and $P(Q)$ is the form factor of a colloidal particle normalized as $P(0) = 1$. At zero scattering angle, the SANS intensity of the star polymer dispersion is written as:

$$I_{\gamma}(Q \to 0) = n[v_{\mathrm{p}} \Delta\rho^{\mathrm{sld}}(\gamma)]^2 S_{\mathrm{sc}}(Q \to 0), \quad (5)$$

where $\gamma$ is the fraction of deuterated component in the solvent. For soft colloidal liquids, $v_{\mathrm{p}}$ contains two parts, the $n$-independent volume of the dry polymer ($v_{\mathrm{polymer}}$) and the $n$-dependent volume of the cavity ($v_{\mathrm{cavity}}$), as illustrated in the inset of Figure 1.[39,40] Here we denote the contrast of the scattering length in the volume of $v_{\mathrm{p}}$ as $b(\gamma)$. It can be found that:

$$b(\gamma) = v_{\mathrm{p}}\Delta\rho^{\mathrm{sld}}(\gamma) = b_{\mathrm{polymer}} - \rho_{\mathrm{sol}}^{\mathrm{sld}}(\gamma)[v_{\mathrm{polymer}} + v_{\mathrm{cavity}}(n) - v_{\mathrm{cavity}}(n)v_{\mathrm{sol}}h(n)], \quad (6)$$

where $b_{\mathrm{polymer}}$ is the scattering length of a star, $\rho_{\mathrm{sol}}^{\mathrm{sld}}$ the sld of the solvent, $v_{\mathrm{sol}}$ the volume of a solvent molecule, and $h$ the number density of the solvent molecule in the cavity. Equation 6 can be justified as follows: $(hv_{\mathrm{cavity}})v_{\mathrm{sol}}$ is the volume occupied by solvent molecules in the cavity, and $\rho_{\mathrm{sol}}^{\mathrm{sld}}(hv_{\mathrm{cavity}})v_{\mathrm{sol}}$ is the contribution of the invasive solvent molecules to the scattering length in $v_{\mathrm{p}}$.[41] Thus, the total contribution of the star polymer and invasive solvent to the scattering length in $v_{\mathrm{p}}$ is $b_{\mathrm{polymer}} + \rho_{\mathrm{sol}}^{\mathrm{sld}}hv_{\mathrm{cavity}}v_{\mathrm{sol}}$. By subtracting the solvent background $\rho_{\mathrm{sol}}^{\mathrm{sld}}v_{\mathrm{p}}$, one finds the contrast of the scattering length in $v_{\mathrm{p}}$ expressed by eq 6.

Combining equations 5 and 6, it is found that:

$$\sqrt{\frac{I_\gamma(Q\to 0)}{n}} = -\rho_{\mathrm{sol}}^{\mathrm{sld}}(\gamma)L(n)\sqrt{S_{\mathrm{sc}}(Q\to 0)} + b_{\mathrm{polymer}}\sqrt{S_{\mathrm{sc}}(Q\to 0)}, \quad (7)$$

where $L(n) = v_{\mathrm{polymer}} + v_{\mathrm{cavity}}(n) - v_{\mathrm{cavity}}(n)v_{\mathrm{sol}}h(n)$. The preceding equation shows the linear relation between $\sqrt{I_\gamma(Q\to 0)/n}$ and $\rho_{\mathrm{sol}}^{\mathrm{sld}}(\gamma)$. By measuring the samples with different $\gamma$ and plotting $\sqrt{I_\gamma(Q\to 0)/n}$ as a function of $\rho_{\mathrm{sol}}^{\mathrm{sld}}(\gamma)$, $S_{\mathrm{sc}}(Q\to 0)$ can be obtained from the vertical intercept. Figure 2 illustrates this method. The adjusted $R^2$ values of all linear fittings shown in Figure 2 are above 0.995, indicating a fairly good linearity between $\sqrt{I_\gamma(Q\to 0)/n}$ and $\rho_{\mathrm{sol}}^{\mathrm{sld}}(\gamma)$. Considering that the form of eq 7 does not change by varying the contrast, we perform the linear extrapolation to $\rho_{\mathrm{sol}}^{\mathrm{sld}}(\gamma) = 0$ to find $S_{\mathrm{sc}}(Q\to 0)$. The experimentally determined $S_{\mathrm{sc}}(Q\to 0)$ for the measured concentrations, from low to high,

are 0.665, 0.235, 0.143, 0.0877, 0.0388 and 0.0184. Then applying eq 2, we obtain the values of $\phi_{\mathrm{eff}}$ to be 0.051, 0.18, 0.25, 0.31, 0.41 and 0.49, respectively.

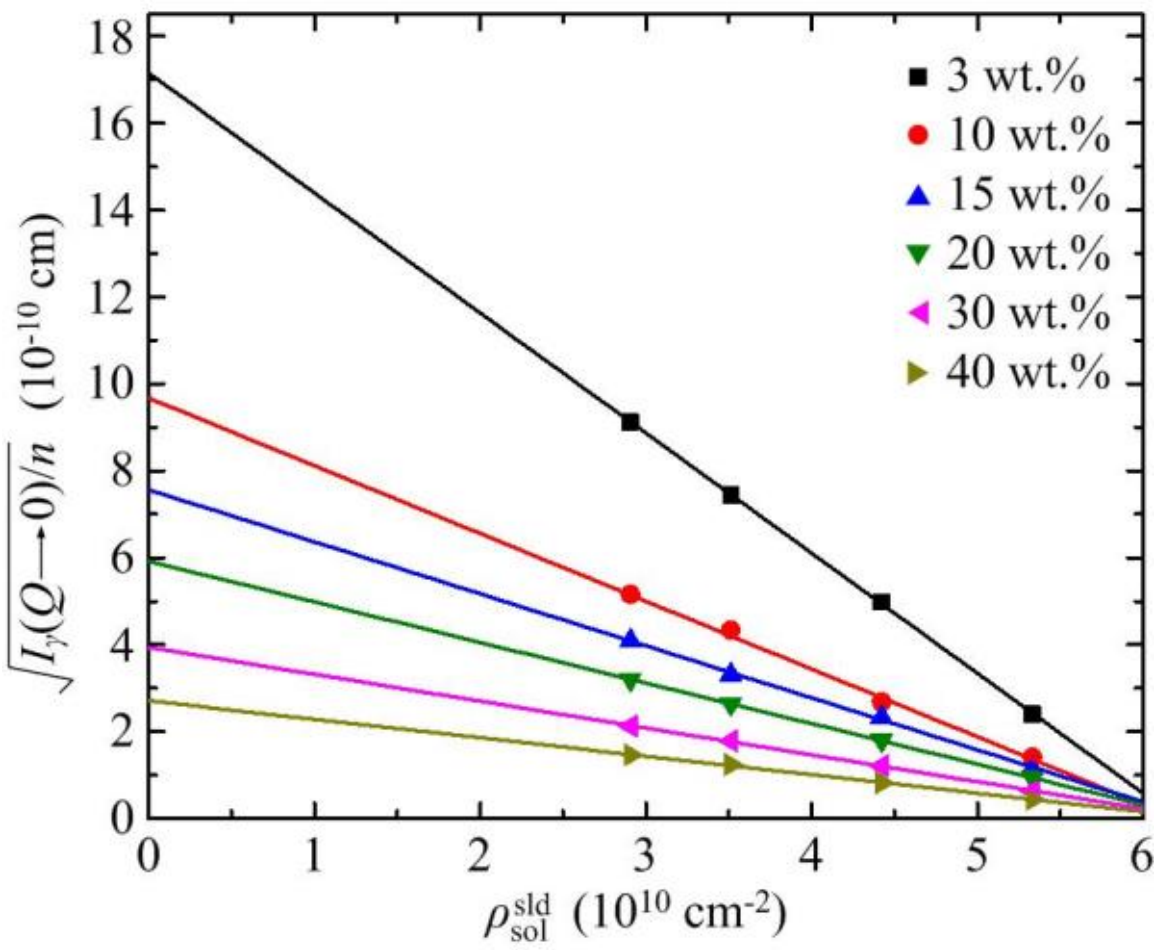

**Figure 2.** Illustration of the contrast-variation SANS method for the determination of $S_{\mathrm{sc}}(Q \to 0)$. In this study, we prepared 4 solvents at $\gamma$ = 40%, 50%, 65% and 80% for each concentration. Deuterated star polymers were used to reduce the incoherent background. The symbols represent the experimental data. The lines are from the linear fitting with eq 7.

Traditionally, people use dynamic light scattering to measure the self-diffusivities of hard-sphere colloids at high $Q$ regions.[8,27] However, it cannot be applied to star polymers, since the local motions, such as the rotational diffusion and intra-molecular motion, are prominent. Here we use the Neutron Spin Echo (NSE) spectrometer with a contrast-match method to measure the short-time self-diffusion of star polymers. We measured the samples at $Q = 0.04$ Å$^{-1}$ to determine $D_{\mathrm{SS}}$. This $Q$ value is much smaller than $2\pi/R_{\mathrm{g}}$ ($R_{\mathrm{g}} =$ 19 Å is the radius of gyration of a star) so that the dynamical contribution of local motions is much less than that of the translational diffusion and thus can be reasonably ignored. In order to obtain the self-dynamics of the star, we prepared the samples composed of a fixed concentration of 3 wt.% protonated star with progressively increasing concentration of fully deuterated stars immersed in solvent whose sld matches that of fully deuterated stars. Because the fraction of protonated stars is relatively dilute, the NSE signal is mainly from the self-motion of a star. The $D_{\mathrm{LS}}$ was measured with the diffusion Nuclear Magnetic

Resonance (NMR) spectroscopy on the same set of samples. Both the measured results of $D_{SS}$ and $D_{LS}$ are shown in Figure 3.[42] It is seen that, as concentration increases, $D_{LS}$ decreases faster than $D_{SS}$.

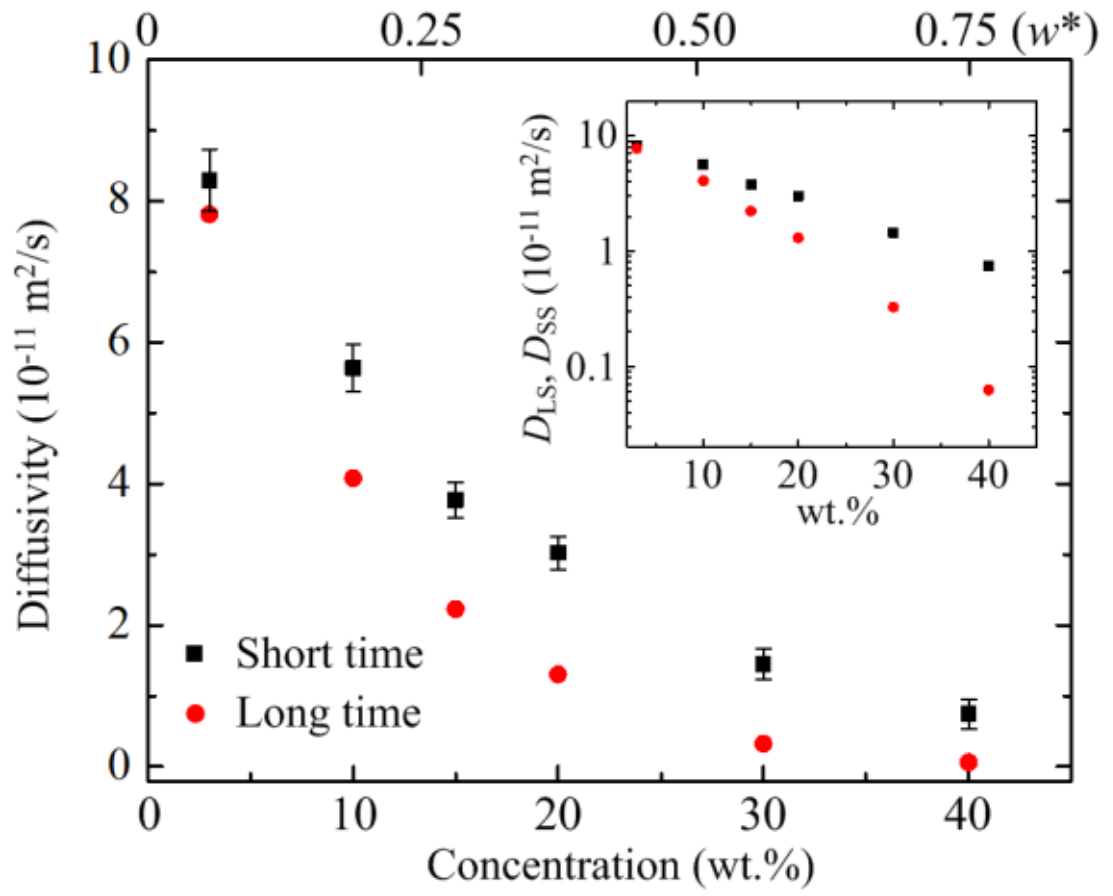

**Figure 3.** The measured short-time and long-time self-diffusivities of the star polymer dispersions as a function of concentration. The upper horizontal scale is in unit of the overlap concentration $w^*$.[43] The inset presents the diffusivities in logarithmic scale.

Knowing $D_{SS}$, $D_{LS}$ and $\phi_{\text{eff}}$, we are able to explore the equivalence in colloidal dynamics. In Figure 4a, we plot $1/D^*$ as a function of $\phi_{\text{eff}}$ for the star polymer dispersion. We also plot $1/D^*$ as a function of the volume fraction for hard spheres. The hard-sphere equation is obtained from the least-square fit of the Brownian dynamics simulation:[44]

$$\frac{1}{D^*} = 1 - 2\phi + 1.272\phi^2 - 1.951\phi^3. \quad (8)$$

It is proven that this form nicely describes the simulations and experimental measurements.[44] From Figure 4a, an excellent agreement between the experimental result and the hard-sphere behavior is seen, which verifies the dynamic equivalence between the soft star polymers and the hard spheres. This observation highlights the effects of excluded volume and overall packing of particles in determining the long-time dynamics of soft colloids. Furthermore, it supports the theoretical conjecture that $D^*$ is not affected by the solvent permeability in soft colloidal suspensions.[44] To make a comparison, we calculate an "apparent volume fraction" of the star, which is defined as $\phi' = 4n\pi R_g^3/3$ and has been used in previous studies of soft particles.[6] The result is presented in the inset of Figure 4a. As seen, the experimental points systematically deviate from the hard-sphere prediction. This deviation suggests that $R_{\mathrm{g}}$ underestimated the excluded volume of soft particles. Similar situation happens to another definition of volume fraction expressed as $\phi'' = n\pi\sigma^{3}/6$,[45] where $\sigma$ is the corona diameter of a star and is related to $R_{\mathrm{g}}$ by $\sigma \approx 1.26R_{\mathrm{g}}$.[22] $\phi''$ is found to be even smaller than $\phi'$ at the same concentration and therefore again underestimates the excluded volume of the star in terms of the long-time self-dynamics. We also evaluate the volume fraction with the hydrodynamic radius $R_\mathrm{h}$ by $\phi_\mathrm{h} = 4n\pi R_\mathrm{h}^3/3$. $R_\mathrm{h}$ is estimated by noting the experimental result that the value of $R_\mathrm{h}$ is close to $\sigma$.[46] The result is shown in the inset of Figure 4a. It is seen that $\phi_\mathrm{h}$ overestimates the excluded volume of the star at concentrations higher than 15 wt.%.

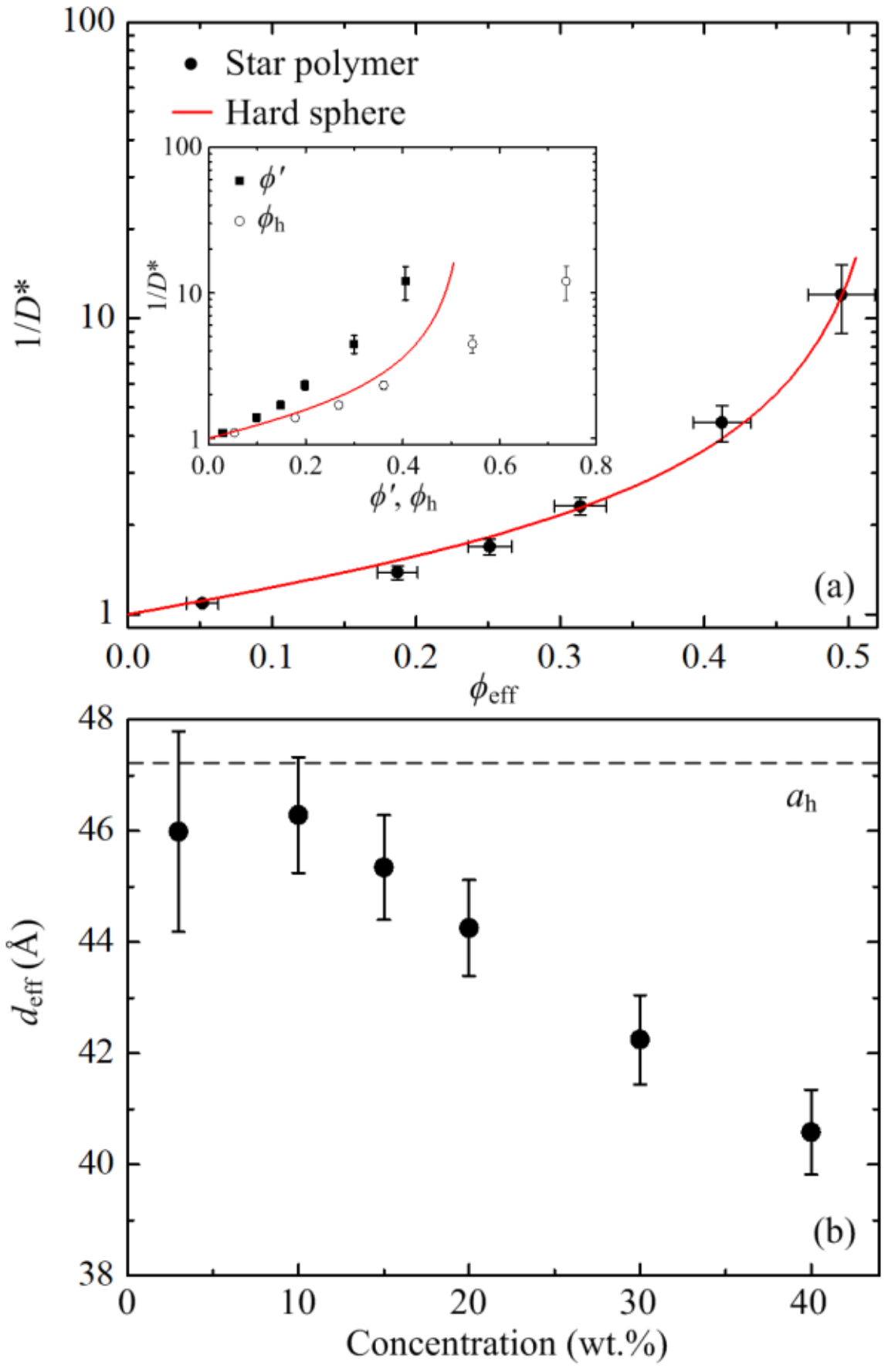

**Figure 4.** (a) $1/D^{*}$ as a function of $\phi_{\mathrm{eff}}$. The symbols represent the experimental result of the star polymer dispersion. The line represents a theoretical calculation for hard spheres.[44] It is seen that the hard-sphere curve gives an excellent prediction to the behavior of the star polymer dispersion. The inset gives the results with two different methods for determining the volume fraction of star polymer: $\phi' = 4n\pi R_{\mathrm{g}}^{3}/3$, $\phi_{\mathrm{h}} = 4n\pi R_{\mathrm{h}}^{3}/3$. (b) Effective diameter of the star $d_{\mathrm{eff}}$ as a function of concentration. $d_{\mathrm{eff}}$ is related to $\phi_{\mathrm{eff}}$ by $\phi_{\mathrm{eff}} = n\pi d_{\mathrm{eff}}^{3}/6$. The dashed line indicates the value of the hydrodynamic diameter $a_{\mathrm{h}}$.

An important observation is the evolution of $d_\mathrm{eff}$ as concentration increases. As shown in Figure 4b, $d_\mathrm{eff}$ is approximately equal to the hydrodynamic diameter $a_\mathrm{h}$ at the lowest concentration. It starts decreasing slightly from a concentration about 15 wt.%, corresponding to the number density of star polymers of $n' = 5.14 \times 10^{18}$ cm$^{-3}$. It is interesting to compare $n'$ with the overlap number density $n^*$ defined as:

$$n^* \approx \phi_\mathrm{RCP}/\tfrac{4}{3}\pi R_\mathrm{g}^3, \quad (9)$$

where $\phi_\mathrm{RCP} = 0.637$ is the random-close-packing volume fraction.[47] Note that, in other studies, $n^*$ is usually defined as $n^* \approx 3/4\pi R_\mathrm{g}^3$.[48,49] These definitions are within the same order. $n^*$ calculated with eq 9 is $2 \times 10^{19}$ cm$^{-3}$, which is much larger than $n'$. The same conclusion can be drawn by calculating $n^*$ using $d_\mathrm{eff}/2$ or $\sigma/2$ instead of $R_\mathrm{g}$. The difference between $n'$ and $n^*$ suggests that the star undergoes conformational change before the physical overlap develops. This result is consistent with our previous study on dendrimer solutions.[50]

Actually, seeking for the analogy between hard-sphere and soft-core liquids has been an essential topic in liquid physics since at least fifty years ago.[51] It was shown that the intermolecular structure of liquids with strongly isotropic repulsive potential can be understood in terms of the packing of equivalent hard-sphere systems.[35] The perturbation theory built on this idea successfully describes the structure and thermodynamics of Lennard-Jones liquids.[52] The hard sphere-soft core analogy has also been observed in dynamic properties in simulations.[9,13,53] Notice that, other dynamic scaling scenarios have been given for temperature-driven liquid-glass transition where hard-sphere properties do not play a substantial role.[54-56] To summary various liquids' universal behaviors, Dyre *et al.* proposed a concept of isomorph for liquids,[57-59] which suggests that the liquids have nearly the same structure and dynamic correlation functions if they are isomorphic. This scaling works very well for liquids with inverse-power-law potentials and also for hard-sphere and Lennard-Jones systems. Our result is generally consistent with this scenario. A special feature of this work is the way to determine $\phi_{\mathrm{eff}}$. Compared with other methods,[10,11,60,61] eq 2 can be directly accessed by SANS measurements for colloidal suspensions without knowing the functional form of interparticle potential. Therefore, it could be of particularly importance for the experimental investigation on colloidal systems.

This work, as well as many previous studies on soft core-hard sphere dynamic analogy, assumes that the dynamics is fully determined by structure. The relevant structural parameter and dynamics evolve monotonically as concentration increases in the studied region. However, the situation at concentrations higher than the overlap concentration could be complicated. Simulation and experiment show that near the jamming point the dynamics and structure of soft-core particles exhibit uncommon discrepancy.[62] Recent study suggests that the crossing of the barriers in energy landscape needs to be considered in finding the effective volume fraction.[63] Moreover, recent simulations found that ultrasoft colloids exhibit reentrant glass transition above the overlap concentration.[64-66] These findings indicate that the strong overlap between soft particles leads to unusual dynamical behaviors.

We notice that computer simulation study reports that the ultrasoft potential developed for star polymers[22] violate the hard sphere-soft core dynamic analogy.[67] The inconsistency between the simulation and our result is probably due to two reasons. First, our samples have a small number of monomers on each arm, which is not perfectly consistent with the physical background of the ultrasoft potential. Second, the theoretical potential form underestimates the excluded volume of the star at finite concentrations. In fact, the SANS spectrum calculated with the ultrasoft potential exhibits a higher $Q_{\mathrm{max}}$ ($Q_{\mathrm{max}}$ is the position of the main peak of $S(Q)$) and a lower intensity compared with the measured spectrum at intermediate concentrations, as shown in Figure 2 of Reference 22. This indicates that the theoretical potential underestimates the strength or the range of exclusion of the interparticle interaction.

In conclusion, we measure star polymer dispersions to study the dynamics of soft colloidal liquids. We find that the slowing of the long-time self-diffusivity, normalized by the short-time self-diffusivity, can be scaled to the hard-sphere behavior with an effective volume fraction determined using the condition of compressibility equivalence. This result provides a new insight in understanding the long-time self-diffusion of soft colloids from a thermodynamic view. In addition, we give a simple method to determine the osmotic compressibility of soft colloidal liquids with SANS.

We are grateful to Dr. J. F. Douglas for valuable discussion. We acknowledge the supports of NIST Center for Neutron Research, Spallation Neutron Source at Oak Ridge National Laboratory, and Institut Laue-Langevin, in providing the neutron research facilities, and NMR facility in University of Arizona in performing the NMR measurement. Access to Neutron Spin Echo and SANS at NIST was provided by the Center for High Resolution Neutron Scattering, a partnership between the National Institute of Standards and Technology and the National Science Foundation under Agreement No. DMR-1508249. This research at SNS of Oak Ridge National Laboratory was sponsored by the Scientific User Facilities Division, Office of Basic Energy Sciences, U.S. Department of Energy. The research at Tsinghua University was supported by the Starting Grant from Tsinghua University.